\documentstyle[prl,aps,twocolumn]{revtex}
\begin{document} 
\draft
\title{Microscopic Model for Sequential Tunneling
in Semiconductor Multiple Quantum Wells.}
\author{ Ram\'on Aguado and Gloria Platero \\}
\address{Instituto de Ciencia de Materiales (CSIC), 
Cantoblanco, 28049 Madrid, Spain.}
\author{Miguel Moscoso and Luis L. Bonilla}
\address{
Escuela Polit\'{e}cnica Superior
Universidad Carlos III de Madrid
Butarque 15, 28911 Legan\'{e}s, Spain}
\maketitle   
\begin{abstract}
\narrowtext
We propose a selfconsistent microscopic model of vertical 
sequential tunneling through a multi-quantum well.
The model includes a detailed description of the contacts, uses the Transfer
Hamiltonian for expressions of the current and it treats the 
Coulomb interaction within a mean field approximation.
We analyze the current density through a double well and a
superlattice and study the formation of electric 
field domains and multistability
coming from the Coulomb interaction. Phase diagrams of 
parameter regions (bias, doping in the heterostructure and in the contacts, 
etc) where the different solutions exist are given.
\end{abstract}


\narrowtext
\par
Coulomb interaction in heterostructures 
with 
large area wells is a small effect compared with the energy 
difference between non-interacting eigenstates of the 
structure. Therefore a mean field model gives, for many 
purposes a good description of the system.
Among features of the transport properties 
having their origin in Coulomb 
interaction, intrinsic bistability has great importance. 
This physical phenomenon arises from the nonlinear effect 
of the electric charge on the induced electrostatic 
potential, and it has been predicted and observed in 
double barrier structures (DB) \cite{1,2,3,4}. 
 Furthermore in the presence of a laser polarized in 
the sample growth direction, new bistability regions have 
been theoretically predicted  \cite{5}.
\noindent In this paper we deal with the statics and dynamics of vertical
transport through biased heterostructures whose main mechanism 
is sequential tunneling. This is a topic which has attracted
a great deal of attention in recent times.
In weakly coupled superlattices, multistability due to domain 
formation has been much studied both theoretically and
experimentally, \cite{6,7,8,8.5}. 
When the charge in the superlattice is small due to
lower doping in the wells, self-sustained current oscillations
and chaos due to domain dynamics are possible 
\cite{9,10,11}. So far, the most successful modeling
of these phenomena use discrete rate equations for the 
electron density and electric field in each well, 
plus constitutive laws for the current, bias, 
boundary and initial conditions, \cite{7,8,12}.
The laws may be phenomenological \cite{8}
or obtained from microscopic considerations, 
\cite{7,13,14}. In all cases cited, the boundary
conditions were selected in a more or less ad hoc manner by 
using the available information from experiments. This is
particularly annoying because the boundary conditions select 
the relevant dynamics of electric field domains in the 
oscillatory regime \cite{15}.
In this paper we present a microscopic model which includes in 
a natural way boundary conditions due to the emitter and 
collector regions of a multiwell structure (MW). We then solve it 
for the cases of a double quantum well (DQW) and a superlattice (SL). 
The presence of intrinsic bistability is demonstrated through
phase diagrams and I-V characteristics obtained 
by numerical simulation and by means of numerical 
continuation of stationary solution branches.
 The main ingredients of our model are as follows: 
 we assume that the characteristic time of 
intersubband relaxation due to scattering (about 0.1 ps 
for optical phonon scattering \cite{16}) is much smaller than the 
tunneling time (less than 0.5 ns), which is in turn much 
smaller than the dielectric relaxation times responsible for 
reaching a steady state (about 10 ns for the 9nm/4nm GaAs/AlAs
superlattices of Ref.~\onlinecite{9}). This separation
of time scales, as well
as the configuration of a typical 
sample allows us to consider that only the ground
state
of each well is populated and that the
tunneling processes are stationary. Our
assumptions then justify using rate equations for the 
electron densities at each well with relations 
for the currents calculated by means of the Transfer Hamiltonian 
(TH) \cite{17}. 
The rate equations for
the electron densities imply that the interwell currents and 
the currents from the emitter and to the collector are all
equal to the total current in the stationary case 
(a form of Amp\`ere's law may be derived in the time-dependent 
case). Furthermore, since no current is created or destroyed 
in the MW, the total charge in it (emitter and 
collector included) is zero. Finally, the electrostatics is
simplified by assuming that the charges are concentrated on 
2D planes located at the wells, emitter and collector regions, 
as indicated in Fig.~\ref{fig1} and further explained below. 

The hamiltonian for 
independent electrons in a MW under dc bias is:
\begin{eqnarray}
H&=&\sum_{k_{i},\, i\in \{L,R\}}E_{k_{i}}{\bf
c^{\dag}_{k_{i}}c_{k_{i}}}
+\sum_{i=1}^{N}\sum_{k_{i}}E_{k_{i}}{\bf
d^{\dag}_{k_{i}}d_{k_{i}}}\nonumber\\
&+&\sum_{k_{i}k_{j}
\left\{ \begin{array}{ll}
i=L & \mbox{j=1} \\
i=R & \mbox{j=N}
\end{array}
\right. }
(T_{k_{i} k_{j}}{\bf c^{\dag}_{k_{i}}d_{k_{j}}}+
H.c)\nonumber\\
&+&\sum_{i=1}^{N-1}\sum_{k_{i} k_{i+1}}(T_{k_{i} k_{i+1}}{\bf
d^{\dag}_{k_{i}}d_{k_{i+1}}}+H.c)
\end{eqnarray}
Here $c^{\dag}_{k_{i}}(c_{k_{i}})$ are the creation (annihilation) 
operators in the leads and $d^{\dag}_{k_{i}}(d_{k_{i}})$  are 
the creation (annihilation) operators in the wells, and $T_{k_{i} k_{j}}$ 
are the tunneling matrix elements. The latter depend on the local electric
field and 
must be calculated self-consistently for each bias. 
Applying the TH under the assumptions listed before, we 
obtain the following expressions for the tunneling currents where 
$J_{e,1}$ and $J_{N,c}$ are the currents in the contacts 
and $J_{i,i+1}$ the interwells currents: 
\begin{eqnarray}
J_{e,1}&\equiv & J_{0,1} = \frac{2ek_{B}T}{\pi^{2}\hbar}\sum_{j=1}^{n}
\int A_{Cj}^{1}(\epsilon)\,
T_{1}(\epsilon)\,\nonumber\\
&\times&\ln \left[\frac{1+e^{\frac{\epsilon_{F}-\epsilon}{k_{B}T}}}
{1+e^{\frac{\epsilon_{\omega_{1}}-\epsilon}{k_{B}T}}}\right]\, 
d\epsilon\nonumber\\
J_{i,i+1}& = &\frac{2e\hbar k_{B}T}{\pi^{2}m^{*}}\sum_{j=1}^{n}
\int A_{C1}^{i}(\epsilon)\, A_{Cj}^{i+1}(\epsilon)\,
T_{i+1}(\epsilon)\,\nonumber\\ 
&\times&\ln \left[\frac{1+e^{\frac{\epsilon_{\omega_{i}}-\epsilon}{k_{B}T}}}
{1+e^{\frac{\epsilon_{\omega_{i+1}}-\epsilon}{k_{B}T}}}
\right]\, d\epsilon,  \nonumber\\
J_{N,c}&\equiv & J_{N,N+1} = \frac{2ek_{B}T}{\pi^{2}\hbar}
\int A_{C1}^{N}(\epsilon)\ T_{N+1}(\epsilon)\,\nonumber\\
&\times&\ln \left[\frac{1+e^{\frac{\epsilon_{\omega_{N}}-\epsilon}{k_{B}T}}}
{1+e^{\frac{\epsilon_{F}-eV-\epsilon}{k_{B}T}}}
\right]\, d\epsilon, \label{THM}
\end{eqnarray}
where $i=1,\ldots,N-1$, $N$ is the number of wells, $n$ is the number of
subbands in each well i  with energies $\epsilon_{Cj}^{i} $ 
(refered with respect to 
the origin
of potential drops), and  
$T_{i}$ are the transmision coefficients through the $i$th barrier. The spectral 
functions of the wells are Lorentzians whose widths correspond to the LO 
phonon lifetimes ($\simeq$ 1-10 meV): 
$A_{Cj}^{i}(\epsilon) = \gamma/[(\epsilon -
\epsilon_{Cj}^{i})^2 +\gamma^2]$ for the $i$th well. 
Of course this model can be improved by calculating microscopically 
the self-energies, which could include other scattering mechanisms (e.g.\ 
interface roughness, impurity effects \cite{14}) or even exchange-correlation 
effects (which affect the electron lifetime in a self-consistent way 
\cite{18}). We have assumed that the electrons in each well are in local
equilibrium with Fermi energies $\epsilon_{\omega_{i}}$ which define the
electronic densities $n_i$. For a given set $\{\epsilon_{\omega_{i}}\}$ the 
densities evolve according to the following rate equations: 
\begin{equation}
\frac{dn_{i}}{dt} = J_{i-1,i}-J_{i,i+1}
\hspace{2cm} i=1,\ldots,N.\label{rate}
\end{equation}
In these equations $J_{i,i+1}\equiv  J_{i,i+1}(\epsilon_{\omega_{i}},
\epsilon_{\omega_{i+1}},\Phi)$,
$J_{e,1}\equiv  J_{e,1}(\epsilon_{\omega_{1}},\Phi)$, and $J_{N,c}\equiv 
J_{N,c}(\epsilon_{\omega_{N}},\Phi)$, where $\Phi$ denotes the set of 
voltage drops through the structure which are calculated as
follows.  
The Poisson equation yields the potential 
drops in the barriers, $V_{i}$, and the wells, $V_{wi}$ (see Fig.~\ref{fig1}):
\begin{eqnarray}
\frac{V_{w_{i}}}{w}&=&\frac{V_{i}}{d}+\frac{
n_{i}(\epsilon_{\omega_{i}})-eN_{D}^{w}}{2\varepsilon}
\label{field.inside1}\\
\frac{V_{i+1}}{d}&=&\frac{V_{i}}{d}+\frac{n_{i}(
\epsilon_{\omega_{i}})-eN_{D}^{w}}{\varepsilon}\,,
\label{field.inside2}
\end{eqnarray}
 where $\varepsilon$ is the GaAs static permittivity,
$n_{i}(\epsilon_{\omega_{i}})$ is the 2D 
(areal) charge density at the $i$th well (to be determined), 
$w$ and $d$ are the well and barrier thickness respectively, 
and 
$N_{D}^{w}$ is the 2D intentional doping at the wells. 
The emitter and collector layers can be described by the following
equations \cite{3}: 
\begin{eqnarray}
\frac{\Delta_{1}}{\delta_{1}} = \frac{eV_{1}}{d}\,,
\quad\quad\quad\quad
\sigma = 2 \varepsilon\,\frac{V_{1}}{d}\simeq
eN(E_{F})\Delta_{1}\delta_{1}\,, \label{emitter}\\
\frac{\Delta_{2}}{e}=\frac{V_{N+1}\delta_{2}}{d}-\frac{1}{2 \varepsilon}
eN_{D}\delta_{2}^{2}\,,\quad\quad\quad\quad
\delta_{3} = \frac{\delta_{2}E_{F}}{\Delta_{2}}\,.\label{collector}
\end{eqnarray}
To write the emitter equations (\ref{emitter}), we assume that there 
are no charges in the emitter barrier. Then the electric field 
across $\delta_{1}$ (see Fig.~\ref{fig1}) is equal to that in the emitter 
barrier. 
Furthermore, the areal charge density required to create this electric field 
is provided by the emitter. $N(E_{F})$ is the density of states at the emitter 
$ E_{F}$. To write the collector equations (\ref{collector}), 
we assume that the region of length $\delta_2$ in the collector
is completely depleted of electrons \cite{3} and 
local charge neutrality in the region of length $\delta_3$ 
between the end of the depletion layer $\delta_2$ and the collector.
In order to close the set of equations we impose global charge conservation 
and that all voltage drops across the different
regions must add up to the applied bias: 
\begin{equation}
\sigma+\sum_{i=1}^{N}(n_{i}(\varepsilon_{\omega_{i}})-eN_{D}^{w})
=eN_{D}(\delta_{2}+{1\over 2}\delta_{3}) \,.\label{charge.conserv}
\end{equation}
\begin{eqnarray}
V = \sum_{i=1}^{N+1}V_{i}+\sum_{i=1}^{N}V_{wi}
+ \frac{\Delta_{1}+\Delta_{2}+E_{F}}{e} .\label{bias}
\end{eqnarray}
Note that the right hand side of Eq.(\ref{charge.conserv}) is the
positive 2D charge density depleted in the collector region. 
Instead of the rate equations (\ref{rate}), we can derive a form of 
Amp\`ere's law which explicitly contains the total current density 
$J(t)$. We differentiate (\ref{field.inside2}) with respect to time 
and eliminate $n_i$ by using (\ref{rate}). The result is
\begin{equation}
{\varepsilon\over d}\frac{dV_{i}}{dt} + J_{i-1,i} = J(t),
\quad\quad\quad i=1,\ldots,N+1 ,\label{ampere}
\end{equation}
where $J(t)$ is the sum of displacement and 
tunneling currents. The time-dependent model consists of the $3N+8$ equations 
(\ref{field.inside1}) - (\ref{ampere}) (the currents
are given by Eqs.~(\ref{THM})), which 
contain the $3N+8$ unknowns $\epsilon_{\omega i}$, $V_{w_{i}}$,
($i=1,\ldots, N$), $V_j$ ($j=1,\ldots,N+1$), $\Delta_1$, 
$\Delta_2$, $\delta_k$ ($k=1,2,3$), $\sigma$, and $J$. 
Thus we have a system of equations which, together with appropriate
initial conditions, determine
completely and self-consistently our problem. The boundary
conditions arise in a natural way. Notice that the charge and
electric field at the boundaries cannot be set prior to the
calculation of the whole structure, which all previous models did 
\cite{7,8,14}.\\ 
In this paper we are interested in analyzing the statics of the model
and the stability of the stationary solutions. One way to
do this is to numerically solve the algebraic-differential system 
(\ref{field.inside1}) - (\ref{ampere}) (plus appropriate initial conditions) 
for each bias until a stationary profile is  
reached. This is rather costly, so that we follow this procedure for 
a given value of the bias and then use a numerical continuation 
method to obtain all stationary solution branches in the 
I--V characteristic diagram. This yields both unstable
and stable solution branches. Direct integration of the stationary
equations [dropping the displacement current in (\ref{ampere})] presents
important problems of numerical convergence to the appropriate solutions
in regions of multistability (see below).

We analyze a DQW (sample a) consisting of 90$\AA$ GaAs wells and 40$\AA$ 
$Ga_{.5}Al_{.5}As$ barriers. The doping at both emitter and collector is 
$ N_{D} = 2\times 10^{18}$cm$^{-3}$, and in the wells it is $N^{w}_{D}=1.5
\times 10^{11}$cm$^{-2}$. The half-width of the well states is 
$\gamma =4$meV in Eqs.~(\ref{THM}) and $T=0$. We do not consider 
the effect of other symmetry points in the conduction band than $\Gamma$.
Fig.~\ref{fig2} shows the DQW I--V characteristic for two different values
of $N^{w}_{D}$. In Fig.~\ref{fig2}a the low bias peak corresponds
to $C1C1$ tunneling ($Ci$ are the conduction subbands ordered starting
from that with lowest energy) between adjacent wells. At 
higher bias multistability of stationary solution branches sets in
(three stable solutions coexist at about 0.44 V).
To understand the difference between these solutions, we have depicted
in Fig.~\ref{fig3} the potential profile of three different solutions 
(two stable, one unstable) corresponding to the same voltage (0.41 V).
In Fig.~\ref{fig3}a, the electrons flow from the emitter to $C1$ in the
first well. Then there is $C1C2$ tunneling to the second well. A stable 
solution with lower current density is shown in Fig.~\ref{fig3}c. There
the $C1$ at the first well is below the bottom of the emitter 
layer, then the electrons flow to $C2$ at the first well instead and
J is smaller. A similar situation occurs at higher bias, 
Figs.~\ref{fig3}d to f. The subbands of the two wells are clearly 
off-resonance for the solution with lowest current, Fig.~\ref{fig3}f.
Notice that the current flowing from the emitter to the structure and
the potential profile are 
quite different for the different solutions of our model. 
This shows that boundary conditions assumed in previous 
publications might constitute gross oversimplifications of the physical 
situation. 
In Fig.~\ref{fig4} we present the regions of multistability
depending on the bias and the dimensionless doping inside the 
wells, $\nu = e w\ N_{D}^{w}/[\varepsilon (\epsilon_{C2}-\epsilon_{C1})]$,
or at the emitter $s = e w^{2} N_{D}/[\varepsilon (\epsilon_{C2}-
\epsilon_{C1})]$. We see that there is a lower and upper limit for
both $\nu$ and $s$ to have bistability. 
Then it is possible to control the presence and 
extent of bistability in a sample by changing the doping in the wells or 
at the contacts and the well widths. 
 
In Fig.~\ref{fig5} we plot the I--V curve of a 
$90\AA$GaAs/$40\AA$Ga$_{.5}$Al$_{.5}$As SL with 11 barriers and
10 wells. Its doping is as in Fig.~\ref{fig2}a. The stable
branches are shown as continuous lines in Fig.~\ref{fig5}.
The inset shows three electric field profiles corresponding to three
different voltages. They show the presence of domains in the SL with a
domain wall which moves one well as we change the bias from one branch 
to the next one. Domain coexistence is also shown in 
the SL electrostatic potential profile; see Fig.~\ref{fig1} for
a fixed bias $V_{2}=0.81$V. 
The first branch in Fig.~\ref{fig5} 
corresponds to $C1C1$ tunneling. As $V$ increases, $C1C2$ 
tunneling becomes possible in part of the structure and we have domain 
formation. This situation confirms the findings with other discrete models
with ad hoc boundary conditions \cite{7,8,8.5,12,13,14}. An interesting 
feature in Fig.~\ref{fig5} is that satellite peaks have a smaller current 
than the $C1C1$ peak. This agrees with the results of Ref.~\onlinecite{14}.
Another interesting feature due to the voltage drop at the contacts is
that the number of branches in the I--V curve is less than the number of
wells. This behaviour can be understood looking at the branch at 1.21 V 
where the low field domain occupies the two wells closer to the emitter. 
$C1C2$ tunneling occurs between all the wells in the branch with $V_{3} = 
1.48$V corresponding to a very intense peak of the current.

In summary, we have proposed and solved a microscopic selfconsistent
 model for the
sequential current through a multiwell structure which includes
the current through the contacts and appropriate
boundary conditions. We have obtained the static I--V curve 
and phase diagrams of a DQW and a SL, which display 
multistability associated to domain formation. 
Exchange-correlation (not included in our model) has been
demonstrated to reduce the bistability in a DB \cite{18}.
Including exchange-correlation effects is the aim
of a future work.

We thank M.~Kindelan and J.\ I\~narrea for fruitful discussions 
 and E.~Doedel for sending us his program of numerical
continuation AUTO. One of us (R.A) acknowledge the Fundaci\'on Universidad 
Carlos III de Madrid for financial support. This work has been supported by 
the CICYT (Spain) under contract MAT 94-0982-c02-02 and by the DGICYT
grant PB94-0375.

\begin{figure}
\caption{Electrostatic potential profile in a SL (sample b)
for $V_{2}=0.81 V$.}
\label{fig1}
\end{figure}

\begin{figure}
\caption{ DQW I--V characteristics (sample a). The continuous 
(dotted) lines correspond to stable (unstable) solution 
branches: 
(a) $N_{D}^{W}=1.5\times 10^{11} cm^{-2}$; and
(b) $N_{D}^{W}=4.31\times 10^{11} cm^{-2}$. The inset magnifies the $C1C1$
resonant peak, showing the region of bistability.}
\label{fig2}
\end{figure}

\begin{figure}
\caption{(a) -- (c) The three stationary potential profiles for the DQW 
structure of Fig.~2a at 0.41 V, ordered from highest to lowest current density. 
(d) -- (f) Same for a bias  $V=0.56$V.
In all cases the emitter Fermi energy and the subband energies are
depicted.}
\label{fig3}
\end{figure}

\begin{figure}
\caption{ Phase diagrams showing the regions of multistability for the DQW 
(sample a): (a) dimensionless well doping $\nu$ versus voltage at $s_{1} 
=1.97$ ($ N_{D} = 2\times 10^{18}$cm$^{-3}$),$\nu_{1}=0.46$ corresponds to the
well doping of Fig.~2b; (b) dimensionless contact doping $s$ versus voltage, at  
$\nu_{1}$. }
\label{fig4}
\end{figure}

\begin{figure}
\caption{ I--V characteristic curve of a SL (sample b). 
 The inset shows the electric field distribution through the
SL for three voltages: $V_1 = .69$ V; $V_2 = .81$V; $V_3 = 1.48$V.}
\label{fig5}
\end{figure}



\begin{references}
\bibitem{1} 
E.~S.~Alves et al., 
Electronic Lett.\ {\bf 24}, 1190 (1988).
\bibitem{2} F.W.\ Sheard and G.A.\ Toombs, Appl. Phys. Lett.,
{\bf 52}, 1228 (1988).
\bibitem{3}  V.J.\ Goldman, D.C.\ Tsui  and J.E.\ Cunningham,
Phys.~Rev.~B {\bf 35}, 9387 (1987); 
Phys.~Rev.~Lett.\ {\bf 58}, 1256 (1987). 
\bibitem{4} T.\ Fiig and A.P.\ Jauho, Surf. Sci, {\bf 267}, 392 (1992);
M.\ Wagner and H.\ Mizuta, Jpn.~J.~Appl.~Phys.\ {\bf 32}, 520 (1993).
\bibitem{5} J.~I\~narrea and G.~Platero, Europhys.\ Lett., {\bf 33} (6),
477 (1996).
\bibitem{6}
H.~T.~Grahn et al., 
Phys.~Rev.~Lett.\ {\bf 67}, 1618 (1991).
\bibitem{7}
F. Prengel, A. Wacker, and E. Sch{\"o}ll, Phys.~Rev.~B {\bf 50},  
1705  (1994).
\bibitem{8}
L.~L.~Bonilla et al.,
Phys. Rev. B {\bf 50}, 8644 (1994).
\bibitem{8.5} A. Wacker et al., 
Phys.\ Rev.\ B {\bf 55}, 2466 (1997).
%
\bibitem{9}
J. Kastrup et al., 
 Phys.\ Rev.\ B {\bf 55}, 2476 (1997).
\bibitem{10}
O. M. Bulashenko and L. L. Bonilla, Phys.~Rev.~B {\bf 52}, 
7849 (1995).
\bibitem{11}
Y.~Zhang et al., Phys.\ Rev.\ Lett.\ {\bf 77}, 3001 (1996).
\bibitem{12}
L.~L.~Bonilla et al., SIAM J.~Appl.~Math.\ {\bf 57}(6), (1997), to appear.
\bibitem{13}
 B. Laikhtman and D. Miller, Phys.~Rev.~B {\bf 48}, 5395 (1993).
\bibitem{14}
A.\ Wacker and A.~P.\ Jauho, Physica Scripta {\bf T69}, 321 (1997).  
\bibitem{15}
F.~J.~Higuera and L.~L.~Bonilla,
Physica D {\bf 57}, 164 (1992).
\bibitem{16} F.~Capasso et al., Appl.\ Phys.\
Lett.\ {\bf 48}, 478 (1986).
\bibitem{17} G.~Platero, L.~Brey and C.~Tejedor,
Phys.~Rev.~B {\bf 40}, 8548 (1989).
\bibitem{18} N.\ Zou et al.,
Phys.~Rev.~B {\bf 49}, 2193 (1994).
\end{references}
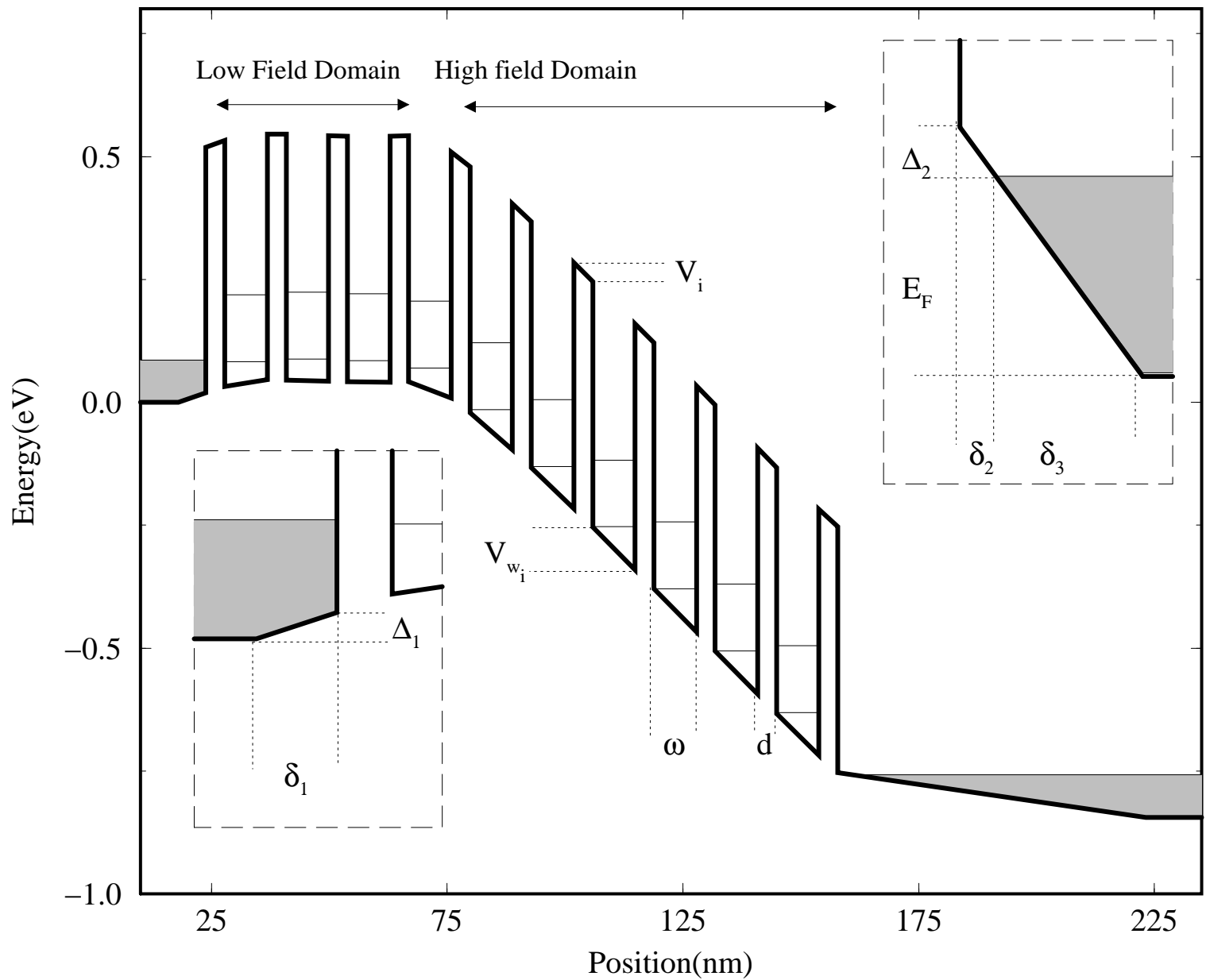
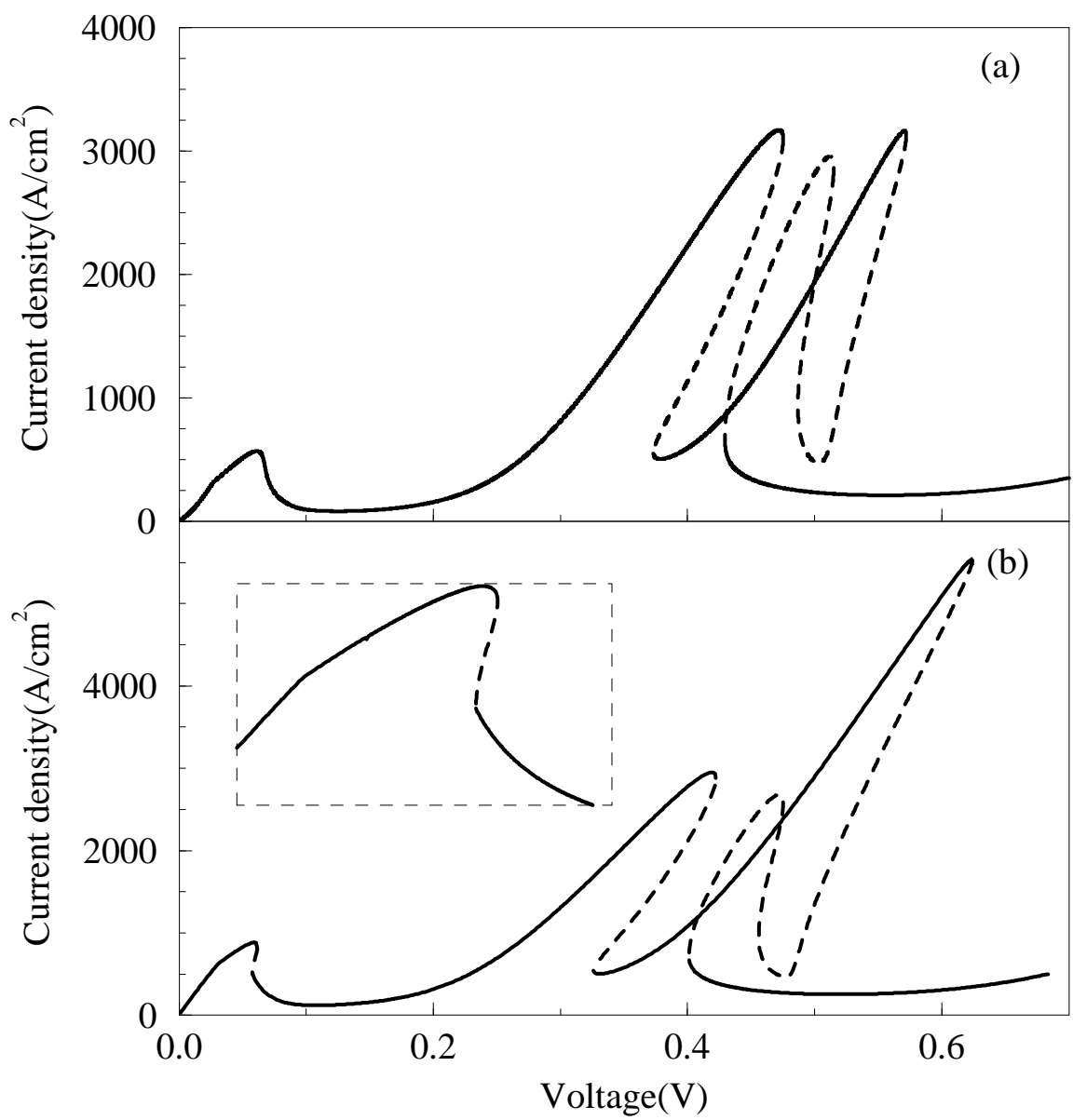
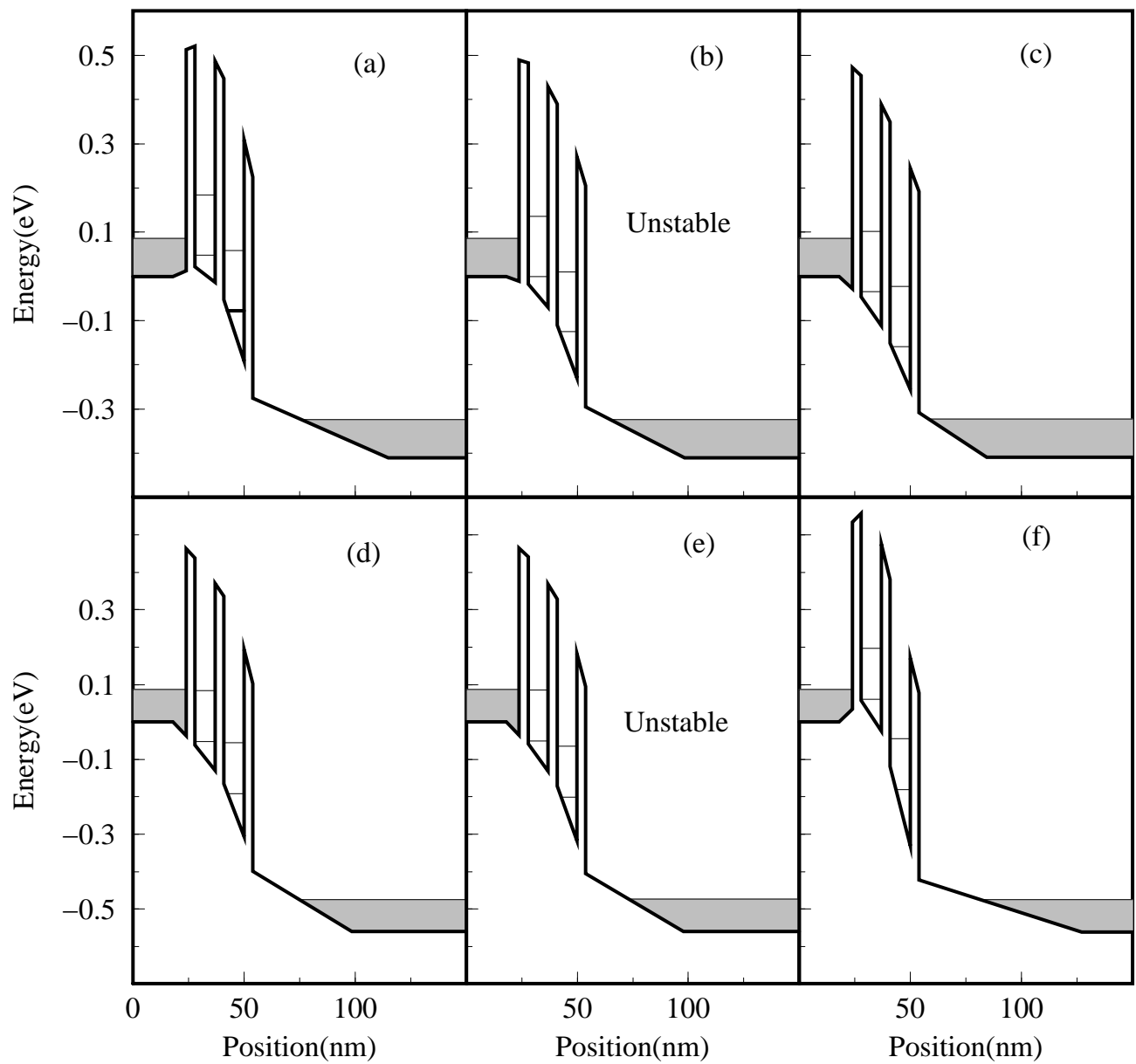
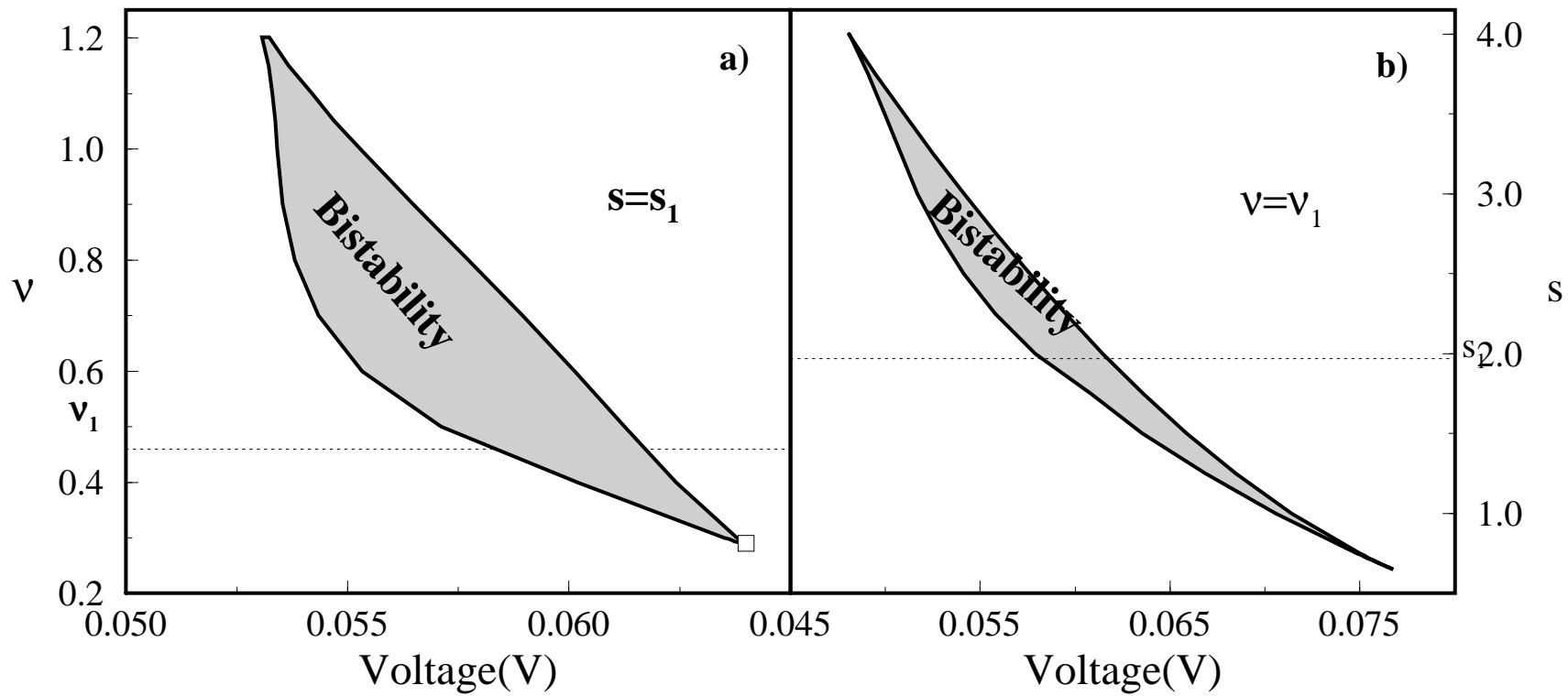
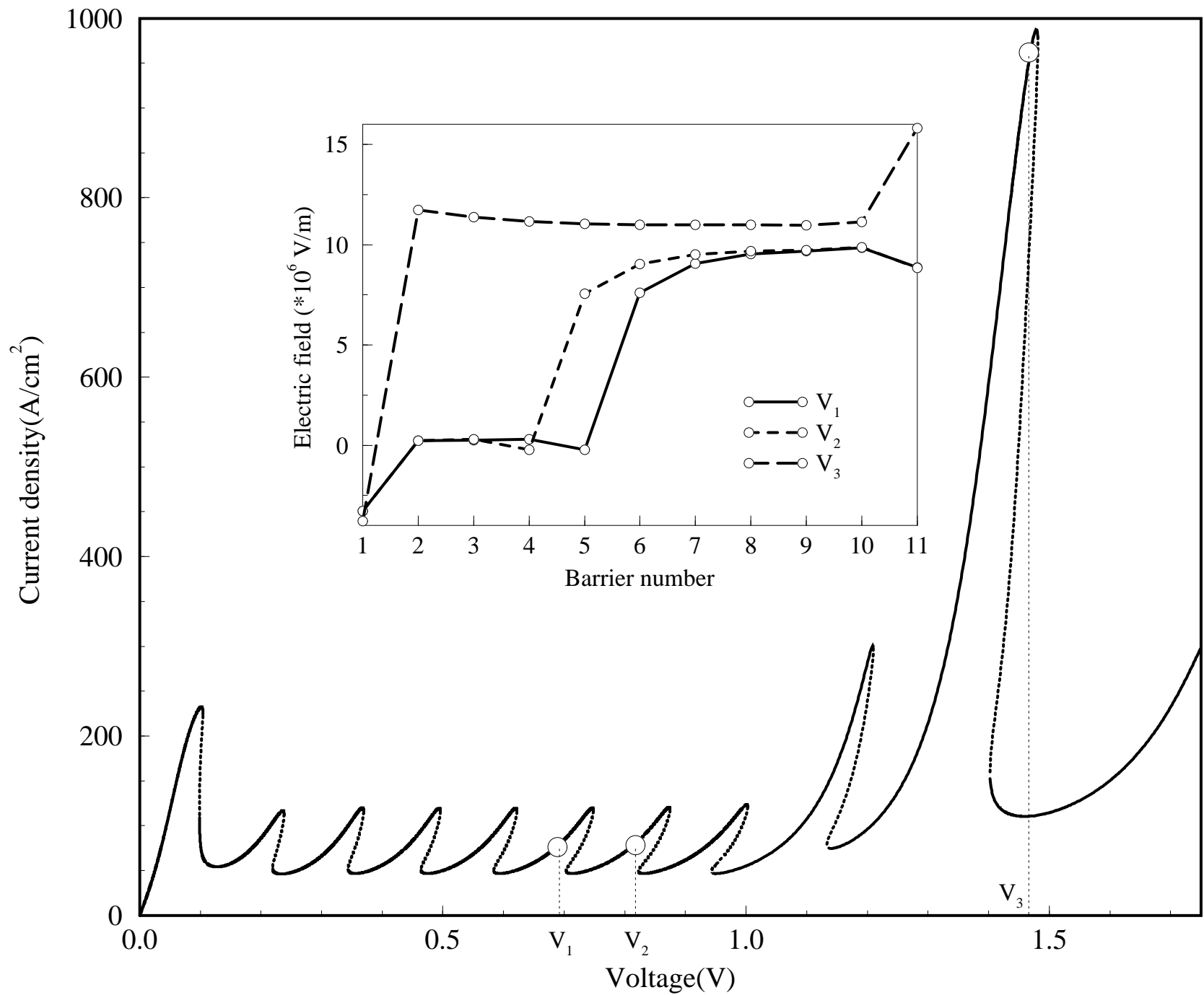
\end{document}